\documentclass[runningheads]{llncs}
\usepackage{lineno}

\usepackage{url}

\usepackage{amsmath}
\usepackage{amssymb}
\usepackage{stmaryrd}
\usepackage{mathtools}
\usepackage{physics}

\usepackage{array}
\usepackage{float}  
\usepackage{caption}

\usepackage{listings}
\usepackage{verbatim}
\usepackage{alltt}
\usepackage{paralist}
\usepackage[normalem]{ulem}

\usepackage{todonotes}

\usepackage[linesnumbered,ruled,vlined]{algorithm2e}

\newcommand\site[1]{\footnote{\url{#1}}}

\usepackage{tikz}
\usetikzlibrary{arrows,positioning,matrix,fit,decorations.text}

\newcommand\eqdef{\stackrel{\mathclap{\normalfont\mbox{def}}}{=}}

\makeatletter
\DeclareTextCommand{\textprime}{\encodingdefault}{%
  \mbox{$\m@th'\kern-\scriptspace$}%
}
\makeatother

\renewcommand{\i}{\textit}  
\newcommand{\s}{\textsf}  
\newcommand{\msf}[1]{\ensuremath{\mathsf{#1}}}
\newcommand{\mi}[1]{\ensuremath{\mathit{#1}}}

\newcommand{\true}{\textsf{true}}

\newcommand{\hash}[1]{\ensuremath{#1^{\#}}}

\newcommand{\List}[1]{\ensuremath{\s{List}[#1]}}
\newcommand{\Set}[1]{\ensuremath{\s{Set}[#1]}}

\newcommand{\Interval}[1]{\ensuremath{\s{Interval}[#1]}}
\newcommand{\FinSup}[2]{\ensuremath{\s{FinSup}[#1,\linebreak[0]#2]}}

\newcommand{\supp}{\msf{supp}}

\newcommand{\ignore}[1]{}

\newcommand{\Script}{\ensuremath{\s{Script}}}
\newcommand{\scriptAddr}{\msf{scriptAddr}}
\newcommand{\keyAddr}{\msf{keyAddr}}

\newcommand{\applyScript}[1]{\ensuremath{\llbracket#1\rrbracket}}

\newcommand{\Input}{\ensuremath{\s{Input}}}
\newcommand{\Output}{\ensuremath{\s{Output}}}
\newcommand{\OutputRef}{\ensuremath{\s{OutputRef}}}
\newcommand{\Signature}{\ensuremath{\s{Signature}}}
\newcommand{\Ledger}{\ensuremath{\s{Ledger}}}

\newcommand{\outputref}{\mi{outputRef}}

\newcommand{\id}{\mi{id}}
\newcommand{\lookupTx}{\msf{lookupTx}}
\newcommand{\getSpent}{\msf{getSpentOutput}}

\newcommand{\Tick}{\ensuremath{\s{Tick}}}
\newcommand{\currentTick}{\msf{currentTick}}

\newcommand{\unspent}{\msf{unspentOutputs}}
\newcommand{\txunspent}{\msf{unspentTxOutputs}}

\newcommand{\utxotx}{\msf{Tx}}
\newcommand{\policiesWithChange}{\msf{policiesWithChange}}
\newcommand{\changedSupply}{\msf{changedSupply}}

\newcommand{\verify}{\msf{verify}}
\newcommand{\TxId}{\ensuremath{\s{TxId}}}
\newcommand{\txId}{\msf{txId}}
\newcommand{\txrefid}{\mi{id}}
\newcommand{\Address}{\ensuremath{\s{Address}}}

\newcommand{\idx}{\mi{index}}
\newcommand{\inputs}{\mi{inputs}}
\newcommand{\outputs}{\mi{outputs}}
\newcommand{\validityInterval}{\mi{validityInterval}}
\newcommand{\scripts}{\mi{scripts}}
\newcommand{\forge}{\mi{forge}}
\newcommand{\sigs}{\mi{sigs}}

\newcommand{\addr}{\mi{addr}}
\newcommand{\key}{\mi{key}}
\newcommand{\val}{\mi{value}}  
\newcommand{\append}{\mathbin{+\!\!+}}

\newcommand{\Quantity}{\ensuremath{\s{Quantity}}}
\newcommand{\AssetName}{\ensuremath{\s{AssetName}}}
\newcommand{\PolicyID}{\ensuremath{\s{PolicyID}}}
\newcommand{\Policy}{\ensuremath{\s{Policy}}}
\newcommand{\AssetID}{\ensuremath{\s{AssetID}}}
\newcommand{\Quantities}{\ensuremath{\s{Quantities}}}

\newcommand\B{\ensuremath{\mathbb{B}}}
\newcommand\N{\ensuremath{\mathbb{N}}}
\newcommand\Z{\ensuremath{\mathbb{Z}}}
\renewcommand\H{\ensuremath{\mathbb{H}}}

\newcommand{\emptymap}{\ensuremath{\{\}}}

\usepackage{etoolbox}

\newcommand{\UTXO}{UTXO}

\newcommand{\UTXOma}{\UTXO$_{\textrm{ma}}$}
\newcommand{\UTXOll}{\UTXO$_{\textrm{ll}}$}

\newcommand{\PublicKey}{\ensuremath{\s{PubKey}}}

\newif\iflong
\longtrue

\newcommand{\longelse}[2]{%
  \iflong
    #1%
  \else
    #2%
  \fi
}

\begin{document}

\title{Babel Fees via Limited Liabilities}

\author{%
  Manuel M. T. Chakravarty\inst{1}
  \and
  Nikos Karayannidis\inst{2}
  \and
  Aggelos Kiayias\inst{3,4}
  \and
  Michael Peyton Jones\inst{5}
  \and
  Polina Vinogradova\inst{6}
} 
\authorrunning{M. M. T. Chakravarty et al.}

\institute{
IOHK, Utrecht, The Netherlands
\email{manuel.chakravarty@iohk.io}
\and
IOHK, Athens, Greece
\email{nikos.karagiannidis@iohk.io}
\and
IOHK, Edinburgh, Scotland
\and
University of Edinburgh, Edinburgh, Scotland \email{akiayias@inf.ed.ac.uk}
\and
IOHK, London, England
\email{michael.peyton-jones@iohk.io}
\and
IOHK, Ottawa, Canada
\email{polina.vinogradova@iohk.io}
}

\maketitle

\begin{abstract}
  Custom currencies (ERC-20) on Ethereum are wildly popular, but they are second class to the primary currency Ether. Custom currencies are more complex and more expensive to handle than the primary currency as their accounting is not natively performed by the underlying ledger, but instead in user-defined contract code. Furthermore, and quite importantly, transaction fees can only be paid in Ether.

In this paper, we focus on being able to pay transaction fees in custom currencies. We achieve this by way of a mechanism permitting \emph{short term liabilities} to pay transaction fees in conjunction with offers of custom currencies to compensate for those liabilities. This enables block producers to accept custom currencies in exchange for settling liabilities of transactions that they process.

  We present formal ledger rules to handle liabilities together with the concept of \emph{babel fees} to pay transaction fees in custom currencies. We also discuss how clients can determine what fees they have to pay, and we present a solution to the knapsack problem variant that block producers have to solve in the presence of babel fees to optimise their profits.
\end{abstract}


\section{Introduction}

Custom currencies, usually following the ERC-20 standard, are one of the most popular smart contracts deployed on the Ethereum blockchain. These currencies are however second class to the primary currency Ether. Custom tokens are not natively traded and accounted for by the Ethereum ledger; instead, part of the logic of an ERC-20 contract replicates this transfer and accounting functionality. The second class nature of custom tokens goes further, though: transaction processing and smart contract execution fees can only be paid in Ether --- even by users who have got custom tokens worth thousands of dollars in their wallets.

The above two limitations and the disadvantages they introduce
seem hard to circumvent.
After all, it seems unavoidable
that custom tokens  must be issued by a smart contract
and interacting with a smart contract requires 
fees in the primary currency. 
Still, recent work addressing the first limitation, 
showed that it can be tackled: by introducing 
\emph{native custom tokens} (see e.g.,~\cite{utxoma}) it is possible to allow custom tokens to reuse the transfer and accounting logic that is already part of the underlying ledger. This is achieved  without the need for a global registry or similar global structure via the concept of \emph{token bundles} in combination with \emph{token policy scripts} that control minting and burning of custom tokens. 
Nevertheless, even with native custom tokens, transaction fees still need to be paid in the primary currency of the underlying ledger.

To the best of our knowledge the only known technique to tackle the second limitation is in the context of Ethereum: the 
  \emph{Ethereum Gas Station Network (GSN)}\footnote{\url{https://docs.opengsn.org/}}. The GSN attempts to work around this inability to pay fees with custom tokens by way of a layer-2 solution, where a network of relay servers accepts fee-less \emph{meta-transactions} off-chain and submits them, with payment, to the Ether\-eum network. In return for this service, the GSN may accept payment in other denominations, such as custom tokens.  Meta-trans\-actions have the downside that in order to remove trust from intermediaries, custom infrastructure in {\em every} smart contract that wants to accept transactions via the GSN is needed. This has the serious downside that
 GSN users are only able to engage with the subset of the ledger state
 that explicitly acknowledges the GSN network. Beyond reducing the scope of GSN transactions, this introduces additional complexity on smart contract development including the fact that participating smart contracts must be pre-loaded with funds to pay the GSN intermediaries for their services. 

Motivated by the above, we describe a solution that lifts this second limitation of custom tokens entirely and without requiring any modification to smart contract design. More specifically, we introduce the concept of \emph{babel fees,} where fee payment is possible in any denomination that another party values sufficiently to pay the actual transaction fee in the primary currency. 
Our requirements for babel fees go beyond what GSN offers and are summarized as follows: (1) participants that create a babel fee transaction should be able to create a normal transaction, which will be included in the ledger exactly as is (i.e., no need for meta-transactions or specially crafted smart contract infrastructure) and (2) the protocol should be non-interactive in the sense that a single message from the creator of a transaction to the participant paying the fee in the primary currency should suffice. In other words, we want transaction creation and submission to be structurally the same for transactions with babel fees as for regular transactions.

Our implementation of babel fees is based on a novel ledger mechanism, which we call \emph{limited liabilities.} These are negative token amounts (debt if you like) of strictly limited lifetime. Due to the limited lifetime of liabilities, we prevent any form of inflation (of the primary currency and of custom tokens).

Transactions paid for with babel fees simply pay their fees with primary currency obtained by way of a liability. This liability is combined with custom tokens offered to any party that is willing to cover the liability in exchange for receiving the custom tokens. In the first instance, this allows block producers to process transactions with babel fees by combining them with a second fee paying transaction that covers the liability and collects the offered custom tokens. More generally, more elaborate matching markets can be set up.

We describe native custom tokens and liabilities in the context of the \UTXO\ 
ledger model. However, our contribution is more general and we sketch in \longelse{Appendix~\ref{sec:appx:account-based-ledgers}}{the unabridged version~\cite[Appendix~C]{longversion}} 
how it can be adapted for an account-based ledger. In summary, this paper makes 
the following contributions:

\begin{itemize}
\item We introduce the concept of \emph{limited liabilities} as a combination of negative values in multi-asset token bundles with batched transaction processing (Section~\ref{sec:overview}).
\item We introduce the concept of \emph{babel fees} on the basis of limited liabilities as a means to pay transaction fees in tokens other than a ledger's primary currency (Section~\ref{sec:overview}).
\item We present formal ledger rules for an \UTXO\ multi-asset ledger with limited liabilities (Section~\ref{sec:ledger-rules}).
\item We present a concrete spot market scheme for block producers to match babel fees (Section~\ref{sec:babel-fees-impl}).
\item We present a solution to the knapsack problem that block producers have to solve to maximise their profit in the presence of babel fees (Section~\ref{sec:tx_selection}).
\end{itemize}
We discuss related work in Section~\ref{sec:related-work}.

\section{Limited liabilities in a Multi-Asset Ledger}
\label{sec:overview}

To realise babel fees by way of liabilities, we require a ledger that supports multiple \emph{native} assets --- i.e., a number of tokens accounted for by the ledger's builtin accounting. Moreover, one of these native tokens is the \emph{primary currency} of the ledger. The primary currency is used to pay transaction fees and may have other administrative functions, such as staking in a proof-of-stake system.

\subsection{Native custom assets}

To illustrate limited liabilities and Babel fees by way of a concrete ledger model, we use the \UTXOma\ ledger model~\cite{utxoma} --- an extension of Bitcoin's \emph{unspent transaction output (\UTXO)} model to natively support multiple assets.\footnote{The \UTXOma\ ledger model is in-production use in the Cardano blockchain.} For reference, we list the definitions of that ledger model in \longelse{Appendix~\ref{sec:utxoma}}{the unabridged paper~\cite[Appendix~A]{longversion}}, with the exception of the ledger rules that we discuss in the following section. To set the stage, we summarise the main points of the ledger model definitions in the following.

We consider a ledger $l$ to be a list of transactions \([t_1, \ldots, t_n]\). Each of these transactions consists of a set of inputs $\i{is}$, a list of outputs $\i{os}$, a validity interval $\i{vi}$, a forge field $\val_{\i{forge}}$, a set of asset policy scripts $\i{ps}$, and a set of signatures $\i{sigs}$. Overall, we have 
\begin{align*}
t = {} 
&(\inputs: \i{is}, 
\outputs: \i{os}, 
\validityInterval: \i{vi}, 
\\
&\forge: \val_{\i{forge}},
\scripts: \i{ps},
\sigs: \i{sigs})
\end{align*}
The inputs refer to outputs of transactions that occur earlier on the ledger --- we say that the inputs \emph{spend} those outputs. The outputs, in turn, are pairs of addresses and values: \((\addr: a, \val: v)\), where $\addr$ is the hash of the public key of the key pair looking that output and $\val$ is the \emph{token bundle} encoding the multi-asset value carried by the output. We don't discuss script-locked outputs in this paper, but they can be added exactly as described in~\cite{eutxoma}.

Token bundles are, in essence, finite maps that map an asset ID to a quantity --- i.e., to how many tokens of that asset are present in the bundle in question. The asset ID itself is a pair of a hash of the policy script defining the asset's monetary policy and a token name, but that level of detail has no relevance to the discussion at hand. Hence, for all examples, we will simply use a finite map of assets or tokens to quantities --- e.g.,  
\(
\{\i{wBTC} \mapsto 0.5, \i{MyCoin} \mapsto 5, \i{nft} \mapsto 1 \}
\)
contains $0.5$ wrapped Bitcoin, five $\i{MyCoin}$ and one $\i{nft}$.

The $\forge$ field in a transaction specifies a token bundle of minted (positive) and burned (negative) tokens. Each asset occurring in the forge field needs to have its associated policy script included in the set of policy scripts $\i{ps}$. Moreover, the $\sigs$ fields contains all signatures signing the transaction. These signatures need to be sufficient to unlock all outputs spent by the transaction's inputs $\i{is}$. Finally, the validity interval specifies a time frame (in an abstract unit of ticks that is dependent on the length of the ledger) in which the transaction may be admitted to the ledger.

We call the set of all outputs that (1) occur in a transaction in ledger $l$ and (2) are not spent by any input of any transaction in $l$ the ledger's \UTXO\ set --- it constitutes the ledger's state.

\subsection{Limited liabilities} \label{sec:lim_liabilities}
In a UTXO, the value for a specific token in a token bundle is always positive. 
In other words, the value component of a \UTXO\ is always a composition of 
assets. It cannot include a debt or liability. We propose to \emph{locally} 
change that.

\subsubsection{Liabilities}
We call a token in a token bundle that has a negative value a \emph{liability.} 
In other words, for a token bundle $\val$ and asset $a$, if \(\val(a) < 0\), 
the bundle $\val$ includes an $a$-liability.

\subsubsection{Transaction batches} \label{sec:tx_batch}

In order to prevent liabilities appearing on the ledger proper, we do not allow the state of a fully valid ledger to contain \UTXO{}s whose value includes a liability. We do, however, permit the addition of multiple transactions \emph{at once} to a valid ledger, as long as the resulting ledger is again fully valid; i.e., it's \UTXO\ set is again free of liabilities. We call a sequence of multiple transactions $\mathit{ts}$, which are being added to a ledger at once, a \emph{transaction batch.} A transaction batch may include transaction outputs with liabilities as long as those liabilities are resolved by subsequent transactions in the same batch.

Consider the following batch of two transactions:
{
\begin{align*}
t_1 = {}
&(\inputs: \i{is}, 
\\
&\outputs: [(\addr: \emptyset, \val: \{ T_1 \mapsto -5, T_2 \mapsto 10 \}), 
\\
&\phantom{\outputs: [}(\addr: \kappa_1, \val: \{ T_1 \mapsto 5 \})],
\\
& \validityInterval: \i{vi}, \forge: 0, \scripts: \{\}, \sigs: \i{sigs})
\\
t_2 = {}
&(\inputs: \{ (\outputref: (t_1, 0), \key: \emptyset), i_{T_1}  \}, 
\\
&\outputs: [(\addr:\kappa_2, \val:\{ T_2 \mapsto 10 \})],
\\
& \validityInterval: \i{vi}, \forge: 0, \scripts: \{\}, \sigs: \i{sigs}')
\end{align*}
}
The first output of transaction $t_1$ may be spend by anybody (\(\addr = \emptyset\)). It contains both a liability of \(-5 T_1\) and an asset of \(10 T_2\). The second transaction $t_2$ spends that single output of $t_1$ and has a second input $i_{T_1}$, which we assume consumes an output containing \(5 T_1\), which is sufficient to cover the liability. 

Overall, we are left with \(5 T_1\) exposed in $t_1$'s second output and locked by $\kappa_1$ as well as \(10 T_2\), which $t_2$ exposes in its single output, locked with the key $\kappa_2$. Both transactions together take a fully valid ledger to a fully valid ledger as the liability is resolved within the transaction batch.

We have these two facts: (a) we have one transaction resolving the liability of another and (b) liabilities are not being permitted in the state of a fully valid ledger. Consequently, transaction batches with internal liabilities are either added to a ledger as a whole or all transactions in the batch are rejected together. This in turn implies that, in a concrete implementation of liabilities in a ledger on a blockchain, the transactions included in one batch always need to go into the same block. A single block, however, may contain several complete batches.

\subsubsection{Pair production}
\label{sec:pair-production}

Liabilities in batches enable us to create transactions that temporarily (i.e., within the batch) inflate the supply of a currency. For example, consider a transaction $t$ with two outputs $o_1$ and $o_2$, where $o_1$ contains \(5000 T\) and $o_2$ contains \(-5000 T\). While value is being preserved, we suddenly do have a huge amount of $T$ at our disposal in $o_1$. In loose association with the somehow related phenomenon in quantum physics, we call this \emph{pair production} --- the creation of balancing positive and negative quantities out of nothing.

As all liabilities are confined to one batch of transactions only, this does not create any risk of inflation on the ledger. However, in some situations, it can still be problematic as it may violate invariants that an asset's policy script tries to enforce. For example, imagine that $T$ is a \emph{role token}~\cite{eutxoma} — i.e., a non-fungible, unique token that we use to represent the capability to engage with a contract. In that case, we surely do not want to support the creation of additional instances of the role token, not even temporarily.

In other words, whether to permit pair production or not depends on the asset policy of the produced token. Hence, we will require in the formal ledger rules, discussed in Section~\ref{sec:ledger-rules}, that transactions producing a token $T$ always engage $T$'s asset policy to validate the legitimacy of the pair production.

\subsection{Babel fees}
\label{sec:babel-fees}

Now, we are finally in a position to explain the concrete mechanism underlying 
babel fees. The basic idea is simple: assume a transaction $t$ that 
attracts a fee of $x$ $C$ (where $C$ is the ledger's primary currency), which 
we would like to pay in custom currency $T$. We add an additional \emph{babel 
fee output} $o_\mathit{babel}$ with a liability to $t$ : $o_\mathit{babel} = \{ 
C 
\mapsto -x, T \mapsto y \}$.
This output indicates that we are willing to pay $y$ $T$ to anybody who pays the $x$ $C$ in return. Hence, anybody who consumes $o_\mathit{babel}$ will receive the $y$ $T$, but will at the same time have to compensate the liability of $-x$ $C$. The two are indivisibly connected through the token bundle. Thus, we may view a token bundle that combines a liability with an asset as a representation of an atomic swap.

The transaction $t$ can, due to the liability, never be included in the ledger all by itself. The liability $-x$ $C$ does, however, make a surplus of $x$ $C$ available inside $t$ to cover $t$'s transaction fees.

To include $t$ in the ledger, we need a counterparty to whom $y$ $T$ is worth at least $x$ $C$. That counterparty batches $t$ with a fee paying transaction $t_\mathit{fee}$ that consumes $o_\mathit{babel}$. In addition, $t_\mathit{fee}$ will have to have another input from which it derives the $x$ $C$ together with its own transaction fee, all out of the counterparty's assets. The transaction $t_\mathit{fee}$ puts the $y$ $T$, by itself, into an unencumbered output for subsequent use by the counterparty. Finally, the counterparty combines $t$ and $t_\mathit{fee}$ into a transaction batch for inclusion into the ledger.

In Section~\ref{sec:babel-fees-impl}, we will outline a scheme based on Babel fees and fee paying transactions, where block producing nodes act as fee paying counterparties for transactions that offer Babel fees in the form of custom tokens that are valuable to those block producer. They do so, on the fly, in the process of block production.

\subsection{Other uses liabilities and liabilities on account-based ledgers}

Due to space constraints, we relegate a discussion of other uses of liabilities to \longelse{Appendix~\ref{sec:appx:other-applications}. Moreover, Appendix~\ref{sec:appx:account-based-ledgers} describes how limited liabilities can be realised on an account-based ledger.}{the unabridged paper~\cite[Appendix~B]{longversion}, which, in Appendix~C, also describes how limited liabilities can be realised on an account-based ledger.}
\section{Formal ledger rules for limited liabilities}
\label{sec:ledger-rules}

In this section, we formalise the concept of limited liabilities by building on the \UTXOma\ ledger; i.e., the \UTXO\ ledger with custom native tokens as introduced in existing work~\cite{utxoma}. To add support for limited liabilities, we modify the ledger rules in three ways:
\begin{enumerate}
\item The original \UTXOma\ rules are defining ledger validity by adding transactions to the ledger one by one. We extend this by including the ability to add transactions in \emph{batches}; i.e., multiple transactions at once.
\item We drop the unconditional per-transaction ban on negative values in transaction outputs and replace it by the weaker requirement that there remain no negative values at the fringe of a batch of transactions. In other words, liabilities are confined to occur inside a batch and are forced to be resolved internally in the batch where they are created.
\item We amend the rules about the use of policy scripts such that the script of a token $T$ is guaranteed to be run in every transaction that increases the \emph{supply} of $T$.
\end{enumerate}
In this context, the supply of a token $T$ in a given transaction $t$ is the amount of $T$ that is available to be locked by outputs of $t$. If that supply is larger than the amount of $T$ that is consumed by all inputs of $t$ taken together, then we regard $t$ as increasing the supply. This may be due to forging $T$ or due to pair production (as discussed in Section~\ref{sec:pair-production}).

\subsection{Validity}
\label{sec:validity}

In the original \UTXOma\ ledger rules, we extend a ledger $l$ with one transaction $t$ at a time. In the \UTXOll\ ledger rules (\UTXOma\ with limited liabilities), we change that to add transactions in a two stage process that supports the addition of batches of transactions $\mathit{ts}$ with internal liabilities:
\begin{enumerate}
\item We modify the definition of the \emph{validity} of a transaction $t$ in a ledger $l$ from \UTXOma, such that it gives us \emph{conditional validity} of $t$ in $l$ for \UTXOll\ as defined in Figure~\ref{fig:conditional-validity}.
\item We define validity of a batch of one or more transactions $\mathit{ts}$ by way of the conditional validity of the individual $t\in\mathit{ts}$ together with the \emph{batch validity} of $\mathit{ts}$ in ledger $l$.
\end{enumerate}
We describe the details of these two stages in the following.

\subsection{Stage 1: conditional validity}
\label{sec:conditional-validity}
\begin{figure}[t]
	\begin{enumerate}

	\item
	\label{rule:slot-in-range}
	\textbf{The current tick is within the validity interval}
	\begin{displaymath}
		\msf{currentTick} \in t.\i{validityInterval}
	\end{displaymath}

	\item
	\label{rule:all-outputs-are-non-negative}
	\sout{\textbf{All outputs have non-negative values}}
	\begin{displaymath}
		\text{\sout{For all \(o \in t.\outputs,\ o.\val \geq 0\)}}
	\end{displaymath}

	\item
	\label{rule:all-inputs-refer-to-unspent-outputs}
	\textbf{All inputs refer to unspent outputs}
	\begin{displaymath}
		\{i.\outputref: i \in t.\inputs \} \subseteq \unspent(l).
	\end{displaymath}

	\item
	\label{rule:value-is-preserved}
	\textbf{Value is preserved}
	\begin{displaymath}
		t.\forge + \sum_{i \in t.\inputs} \getSpent(i, l) = \sum_{o \in
		t.\outputs} o.\val
	\end{displaymath}

		\item
		\label{rule:no-double-spending}
		\textbf{No output is locally double spent}
		\begin{displaymath}
			\textrm{If } i_1, i \in t.\inputs \textrm{ and }  i_1.\outputref =
			i.\outputref
			\textrm{ then } i_1 = i.
		\end{displaymath}

		\item
		\label{rule:all-inputs-validate}
		\textbf{All inputs validate}
		\begin{align*}
			\textrm{For all } i \in t.\inputs,\
			\text{there exists }\i{sig}\in t.sigs,
			\verify(i.\key, \i{sig}, \txId(t))
		\end{align*}

		\item
		\label{rule:validator-scripts-hash}
		\textbf{Validator scripts match output addresses}
		\begin{displaymath}
			\textrm{For all } i \in t.\inputs,\ \keyAddr(i.\key) = \getSpent(i,
			l).\addr
		\end{displaymath}

		\item
		\label{rule:forging}
		\textbf{Forging}\\
		$\blacklozenge$~A transaction which changes the supply ---i.e.,
		\(\changedSupply(t, l)\neq\{\}\)---  is only valid if either:
		\begin{enumerate}
			\item the ledger $l$ is empty (that is, if it is the initial
			transaction).
			\item \label{rule:custom-forge}
			$\blacklozenge$~for every policy ID $h \in \changedSupply(t, l)$,
			there
			exists $s \in t.\scripts$ with
			$h = \scriptAddr(s)$.
		\end{enumerate}
		\medskip

		\item
		\label{rule:all-mps-validate}
		\textbf{All scripts validate}
		\begin{align*}
			&\textrm{For all } s \in t.\scripts,\ \\
			&~~
			\applyScript{s}(\scriptAddr(s), t,
			\{\getSpent(i, l) ~\vert~ i~\in~ t.\inputs\}) = \true
		\end{align*}

	\end{enumerate}
	\caption{Conditional validity of a transaction $t$ in a ledger $l$
	permitting liabilities}
	\label{fig:conditional-validity}
\end{figure}
\begin{figure}[h]
	\begin{displaymath}
		\begin{array}{l}
			\textbf{-- output references provided by a transaction} \\
			\multicolumn{1}{l}{\txunspent : \utxotx \rightarrow
			\Set{\s{OutputRef}}}
			\\
			\txunspent(t) ~=~
			{
				\{(\txId(t),1), \ldots, (\txId(id),\left|t.outputs\right|)\}
			}\\
			\\
			\textbf{-- a ledger's \UTXO\ set} \\
			\multicolumn{1}{l}{\unspent : \s{Ledger} \rightarrow
			\Set{\s{OutputRef}}}
			\\
			\unspent([]) ~=~ \emptymap \\
			\unspent(t::l) ~=~
			{
				(\unspent(l) \setminus t.\inputs) \cup \txunspent(t)
			}\\
			\\
			\textbf{-- the outputs spent by the given set of transaction
			inputs} \\
			\multicolumn{1}{l}{\getSpent : \s{Input} \times \s{Ledger}
			\rightarrow \s{Output}}
			\\
			\getSpent(i,l) ~=~
			{l}{
				\lookupTx(l, i.\outputref.\id).\outputs[i.\outputref.\idx]
			}\\
			\\
			\textbf{-- policy IDs of assets whose amount varies} \\
			\multicolumn{1}{l}{\policiesWithChange : Quantities \times
			Quantities \rightarrow \Set{\PolicyID}}
			\\
			\policiesWithChange(\mathit{val}_1, \mathit{val}_2) ~=~
			{
				\{ a.pid \mid a\in\supp(\mathit{val}_1 - \mathit{val}_2)\}
			}\\
			\\
			\textbf{-- policy IDs whose supply changed in the
			transaction}\\
			\multicolumn{1}{l}{\changedSupply : \utxotx \times \s{Ledger}
			\rightarrow \Set{\PolicyID}}
			\\
			\changedSupply(t, l)~=~ \\
			~~~~~~{
				\policiesWithChange(\sum_{o\in\getSpent(t.inputs)} o.value^+,
				\sum_{o\in t.outputs} o.value^+) \cup {}
			}\\
			~~~~~~{
				\policiesWithChange(\sum_{o\in\getSpent(t.inputs)} o.value^-,
				\sum_{o\in t.outputs} o.value^-)
			}\\
			~~~~~~{
				\mathbf{where}
			}\\
			~~~~~~{
				\qquad \val^+(a) = \textbf{if}~\val(a) >
				0~\textbf{then}~\val(a)~\textbf{else}~0
			}\\
			~~~~~~{
				\qquad \val^-(a) = \textbf{if}~\val(a) <
				0~\textbf{then}~\val(a)~\textbf{else}~0
			}
		\end{array}
	\end{displaymath}
	\caption{Auxiliary validation functions}
	\label{fig:validation-functions}
\end{figure}

Conditional validity in \UTXOll\ is defined very much like full validity in \UTXOma. Figure~\ref{fig:conditional-validity} defines the conditions for transactions and ledgers to be conditionally valid, which are mutually dependent. The definitions in Figures~\ref{fig:conditional-validity} and~\ref{fig:validation-functions} are based on the ledger formalisation introduced for \UTXOma~\cite{utxoma}. We do not repeat this formalisation here to favour conciseness, but summarise it in \longelse{Appendix~\ref{sec:utxoma} for ease of reference}{the unabridged paper~\cite[Appendix~A]{longversion}}.
\begin{definition}[Conditional validity of transactions and ledgers]
  A transaction \(t\in\utxotx\) is \emph{conditionally valid} for a conditionally valid ledger \(l\in\s{Ledger}\) during tick \currentTick\ if $t$ abides by the conditional validity rules of Figure~\ref{fig:conditional-validity}, using the auxiliary functions summarised in Figure~\ref{fig:validation-functions}.

  A ledger \(l\in\s{Ledger}\), in turn, is conditionally valid if either $l$ is empty or $l$ is of the form $t::l^{\prime}$ with $l^{\prime}$ being a conditionally valid ledger and $t$ being conditionally valid for $l^{\prime}$.
\end{definition}

Figure~\ref{fig:conditional-validity} highlights the two changes that we are making to the \UTXOma\ rules: firstly, we struck out Rule~(\ref{rule:all-outputs-are-non-negative}), and secondly, we changed Rule~(\ref{rule:forging}) in two places marked with $\blacklozenge$. The removal of Rule~(\ref{rule:all-outputs-are-non-negative}) permits liabilities in the first place. Outputs may now contain negative values and, if they do, the associated transaction is merely conditionally valid. Full validity is now conditional on resolving all liabilities from other transactions that are added in the same batch.

Moreover, the change to Rule~(\ref{rule:forging}) ensures that transactions that change the supply of a token under a policy $s$ with script address $h$ do run the policy script $s$, regardless of whether the change in supply is due to a non-empty forge field \(t.\forge\) or due to pair production. In either case, the script is guaranteed an opportunity to validate that the increase in supply abides by the rules enforced by the token policy. In other words, transactions that contain supply changes that violate the associated token policy are guaranteed to be rejected.

\subsubsection{Changed supply.}
The change in supply is computed with the help of the function \(\changedSupply(t, l)\) (defined in Figure~\ref{fig:validation-functions}) that, for a given ledger $l$, determines all policy script hashes $h$ that control an asset whose supply is changed by the transaction $t$. Such a change may be due to the minting or burning of assets in the transactions forge field $t.\forge$ or it may be due to pair production, as discussed in Section~\ref{sec:pair-production}. The function $\changedSupply$ spots supply changes by comparing the quantity of assets and asset liabilities in the inputs and outputs of a transaction. It uses the helper functions $\val^+$ and $\val^-$ to filter all positive (assets) and negative (liabilities), respectively, out of a token bundle.

\subsubsection{Script validation.}
Rule~(\ref{rule:forging}) uses the set of hashes of policy scripts computed by $\changedSupply$ to check that all the corresponding scripts are included in the $t.\scripts$ field. The scripts in $t.\scripts$ are exactly those that Rule~(\ref{rule:all-mps-validate}) executes.

Note that the primary currency of the ledger may require a special case in this rule. The total supply of the primary currency
may be constant as part of the ledger implementation, and therefore its minting policy will always fail to validate, even
in the case of producing and consuming transient debt. 
This may be addressed in (among others) one of the following ways:
either modify the policy to specifically allow pair production of the primary currency, or modify this rule to not check the primary currency policy at all.

\subsection{Stage 2: batch validity}

For a ledger to be valid, we require that it is conditionally valid and that its state (i.e., the set of unspent outputs) does not contain any negative quantities.
\begin{definition}[Ledger validity]
  A ledger \(l : \s{Ledger}\) is \emph{(fully) valid} if $l$ is conditionally 
  valid and also, $\text{for all, } o\in\unspent(l), o.\val \geq 0.$
\end{definition}

On that basis, we define the validity of a batch of transactions $\mi{ts}$ for a valid ledger $l$.
\begin{definition}[Validity of a batch of transactions]
  A batch of transactions \(\mi{ts} : \List{\utxotx}\) is \emph{(fully) valid} for a valid ledger \(l : \s{Ledger}\) if \(\mi{ts} \append l\) is a fully valid ledger.
\end{definition}

%

\section{Implementing Babel Fees}\label{sec:impl_bfees}
\label{sec:babel-fees-impl}

In this section, we describe a concrete spot market, 
where users can exchange custom tokens via the babel fees mechanism described 
in Section~\ref{sec:babel-fees}. This spot market comprises a set of \emph{sellers} 
$\mathbb{S} = \{s_1,s_2,...,s_n\}$ and a set of \emph{buyers}\footnote{Buyers 
in this market are the \emph{block issuers} of the blockchain.} 
$\mathbb{B} = 
\{b_1,b_2, ...,b_m\}$. Sellers sell bundles of custom tokens to buyers, who in 
return provide primary tokens to cover the fees incurred by the transactions 
submitted by the sellers to the network.

\subsection{Babel offers}

In this context, a transaction with a babel fee output (as per Section~\ref{sec:babel-fees}) essentially constitutes an offer --- specifically, the offer to obtain a specified 
amount of custom tokens by paying the liability in primary tokens included in 
the babel fee output. We define such offers as follows.
\begin{definition} \label{def:batch_offer}
	We define a \emph{babel offer} to be a tuple of the form: 
	\begin{align*}
		\mathit{BabelOffer} \eqdef (\mathit{Tx_{id}}, \mathit{TName}, 
		\mathit{TAmount}, 
		\mathit{Liability})
	\end{align*}
	where $\mathit{Tx_{id}}$ is a unique identifier of the transaction containing the babel fee output,  
	$\mathit{TName}$ is a string corresponding 
	to the name of a custom token, $\mathit{TAmount}$ is a positive 
	integer 
	$\in \mathbb{Z^+}$ corresponding to the amount of tokens offered and 
	$\mathit{Liability}$ is a negative integer $\in \mathbb{Z^-}$ 
	corresponding 
	to the amount in primary tokens that has to be paid for obtaining the 
	tokens.
\end{definition}
Sellers produce such babel offers, which are then published to the network and 
are visible to all buyers.

\subsection{Exchange rates}

In our model, we assume that the spot market of babel offers operates in distinct 
rounds\footnote{In practice this can be the block-issuing rounds.}. In 
every round, a buyer is selected from the set $\mathbb{B}$ at random. The selected 
buyer has the opportunity to accept some of the outstanding offers by paying the 
corresponding 
liabilities. The rational buyer chooses the offers that maximise her utility 
function, which we elaborate in Section~\ref{sec:tx_selection}.

In order to help sellers to make attractive offers, we assume that every buyer 
$i$, $i = 1,2,...,m$ publishes a list 
$L_i[(T_j, \mathit{XR}_j)]$ of exchange 
rates $\mathit{XR}_j$ for every exchangeable custom token $T_j$, $j = 1,...,k$. The list of exchange rates from all buyers $BL[i] = L_i$, $i = 
1,...,m$ is available to all sellers $s \in \mathbb{S}$.
Note that the buyer can set $\mathit{XR}_j=+\infty$ if they don't accept the token. 

Given a specific babel offer $g = (t_g, (\mathtt{token_A}, 
\mathtt{amount_A}, \mathtt{liability_A}))$ offering an amount of a  
custom token 
$\mathtt{token_A}$, and assuming that there is only a single buyer $b$ with a published 
exchange rate for $\mathtt{token_A}$ equal to $\mathit{XR}_A = \frac{\mathtt{token_A}}{\mathtt{primary\ token}}$, an attractive offer 
should adhere the inequality: $\mathtt{amount_A} \geq |\mathtt{liability_A}| 
\mathit{XR}_A$.
%
%
Naturally, an offer gets more attractive to the degree that excess tokens are offered over the minimum needed to meet the exchange rate for the liability. 

\subsection{Coverage}

To generalise to the case where $m$ possible buyers express an interest in $\mathtt{token_A}$, we need to consider the following question: how many $\mathtt{token_A}$ does a seller need to offer to ensure that $P\%$ buyers consider the offer attractive? 

The seller has to choose the cheapest $P_{\mathit{th}}$ percentile from the available exchange rates listed for $\mathtt{token_A}$, which by definition is satisfied by an effective exchange rate that is greater than $P\%$ of the published exchange rates. In other words, for the offer $g$ from above to be attractive to $P\%$ of buyers, the seller needs to choose the amount for $\mathtt{token_A}$ as follows:
\begin{align} \label{eq:comm-attractive}
	\mathtt{amount_A} \geq |\mathtt{liability_A}| \mathit{percentile}(P, \mathtt{token_A}, 
	BL)
\end{align}
where \(\mathit{percentile}(P, \mathtt{token_A}, BL)\) is the lowest exchange rate for $\mathtt{token_A}$, such that it is still greater than $P\%$ of the exchange rates listed for that token in the exchange rate table $BL$. In this case, we say that the offer $g$ has $P\%$ \emph{coverage.}

For example, assume a liability of $0.16$ primary tokens and a set of 
$10$ buyers 
with the following published exchange rates for $\mathtt{token_A}$, $BL_{\mathtt{token_A}} = \{1.63, 
1.38, 3.00, 1.78, 2.00, 1.81\}$. If a seller wants to ensure that more than 
$70\%$ of 
the buyers will consider her offer, she computes the $70$th 
percentile of the exchange rates, which is $2.00$. Thus, the seller 
knows that she has to offer at least $0.16 \times 2.00 = 0.32$ of $\mathtt{token_A}$.

\subsection{Liveness}

Consider a babel offer that is published to the network and assume that 
there is \emph{at least one party} $b_i$ (buyer) that is attracted by this 
offer. The 
interested party will then create a transaction batch $t_{xb}$ (see Section~
\ref{sec:tx_batch}) that covers the 
liability and will publish it to the network with the expectation that this 
will (eventually) be included into a block and be published in the ledger 
implemented by the blockchain. Therefore, it is crucial to ensure 
\emph{censorship resilience} for our Babel offers and show that our spot market 
for Babel offers 
enjoys the property of \emph{liveness}~\cite{backbone}.

If $b_i$ is selected as a block issuer, then she will include the transaction 
batch in the block she will create and thus liveness is preserved. However, if 
$b_i$ 
is never selected as a block issuer (or is selected with a very low 
probability), then we must ensure the accepted offer will eventually be 
included into the blockchain. In the following analysis, we distinguish between 
two cases: a) The case where all buyers are acting rationally (but not 
maliciously) and b) the case 
where a percent of the buyers are controlled by a malicious adversary party. 
Our detailed analysis is presented in \longelse{Appendix~\ref{sec:appx:liveness}}{the paper's unabridged version~\cite[Appendix~D]{longversion}} and has 
shown that our spot market 
indeed enjoys liveness, if the buyers are rational players trying to maximize 
their profit. Moreover, in the case of adversary players, if honest majority 
holds and a Babel offer attracts at least one honest player, then the accepted 
offer will be (eventually) published in the blockchain and thus liveness is 
preserved.
\section{Transaction Selection for Block Issuers}\label{sec:tx_selection}

A block issuer constructs a block of transactions by choosing from a set of 
available transactions called the \emph{mempool}. A rational block issuer tries 
to maximize her utility. In our case, we assume that this utility is a value, 
corresponding to the amount of primary currency earned by this block. These 
earnings come from the transaction fees paid either in primary currency or 
custom tokens. Hence we assume the existence of a utility function of the 
form: $\mathit{utility} :: CandidateBlock \rightarrow Value$, where 
$\mathit{CandidateBlock}$ is 
a list of transactions $\mathit{CandidateBlock} \eqdef 
List[CandidateTransaction]$ and 
$\mathit{Value}$ is an amount $\in \mathbb{Z}^+$ of primary currency at the 
lowest 
denomination.

\subsection{The value of Babel offers}

A candidate transaction residing in the mempool and waiting to be included in a 
block can be either a (single) transaction or a transaction batch
(see Section~\ref{sec:tx_batch}). 
In the following, we define the concept of a \emph{candidate transaction}: 
\begin{definition}\label{def:candidatetx}
	A candidate transaction residing in the mempool is defined as quadruple:
	\begin{align*}
		\mathit{CandidateTransaction} \eqdef (\mathit{Tx_{id}}, \mathit{Value}, 
		\mathit{Liability}, \mathit{Size})
	\end{align*}
    where $\mathit{Tx_{id}}$ is a unique identifier of a transaction (or a 
    transaction batch) in the mempool, $\mathit{Value}$ for the case 
    of transactions 
    corresponds to the transaction fees expressed in the 
    primary currency, while for the case of transaction batches, it 
    corresponds to the total value of the 
    obtained custom tokens expressed as an amount in the primary currency. In 
    the case of transaction batches, $\mathit{Liability} \in \mathbb{Z}^-$ 
    is the amount 
    expressed in the primary currency that has 
    to be paid for covering this liability. In the case of transactions, it    
    equals zero. 
    Finally, $\mathit{Size}$ is the total size of the 
    transaction, or the transaction batch as a whole, expressed in bytes.
\end{definition}

We assume the existence of a function that can transform a Babel 
offer 
(Definition~\ref{def:batch_offer}) into a candidate transaction batch: 
$\mathit{batchVal} :: \mathit{BabelOffer} \rightarrow 
\mathit{CandidateTransaction}$.
%
%
We need this function in order to be able to express the value of the obtained 
custom tokens in primary currency, so that Babel offers are comparable 
to the transaction fees of conventional transactions. Any such conversion 
function might 
be chosen by the block issuer based on her business logic of how to evaluate a 
specific offer. In particular, one reasonable approach to defining the conversion function is the following:
\begin{align*}
	{\mathit{Value}}={} &~~ \sum_{\forall \mathit{token} \in \mathit{BabelOffer}}
	\mathit{TAmount}  
	\frac{\mathit{nominalVal}}
	{\abs{\mathit{Liability\ per\ token}}} 
	\mathit{nominalVal} 
	\\
	{\phantom{\mathit{Value}}}={} &~~ \sum_{\forall \mathit{token} \in \mathit{BabelOffer}}
	\frac{(\mathit{TAmount} \times \mathit{nominalVal})^2} 
	{\abs{\mathit{Liability}}} 
\end{align*}
The nominal value of the token, $\mathit{nominalVal}$, is essentially the 
current rate $\frac{primary\ currency}{custom\ token}$; i.e., it expresses what 
amount 
of primary currency one custom token is worth. Therefore, if the exchange rate 
between 
a custom token $T$ and the primary currency $A$ is $3:1$, then 
$\mathtt{nominalVal} = 0.33 A$. Of 
course, this rate is dynamic and 
it is determined 
by market forces just like with fiat currencies and Bitcoin fees. We assume 
that this information is available to the block producer, when they need to 
select 
candidate transactions from the mempool to include in a new block. 
In fact, block issuers can publish exchange rates for specific tokens they 
consider acceptable (as discussed in see Section~\ref{sec:impl_bfees}). 
Intuitively, the higher 
the nominal value, the more 
valuable the token is to the block issuer.

Hence, whenever a block issuer tries to assemble a block they face the 
following optimization problem:
\begin{definition}
	The \emph{transaction selection problem} $TxSelection(n,S_B,M, 
	R)$
	is the problem of filling a candidate block of 
	size $S_B$, with a subset $B_n \subseteq M$ of $n$ available 
	candidate transactions $M = \{tx_1, tx_2, ...,tx_n\}$, where we use $B_n 
	\subseteq \{1,2,..,n\}$, without spending more than a	reserve $R$ of 
	available primary currency on 
	liabilities, in such a 
	way 
	that $\mathit{utility(B_n)} \geq \mathit{utility(B'_n)}$ $\forall$ block 
	$B'_n \subseteq M$. 
	Every candidate transaction $tx_i = (i, v_i, l_i, s_i)$, for $i = 1,...,n$ 
	is defined according to Definition \ref{def:candidatetx} and has a fixed 
	liability $l_i$ and size $s_i$ in bytes. We assume that the value of 
	a 	candidate transaction that corresponds to a Babel offer is 
	not fixed; instead, it decreases (just as its desirability) as 
	we select candidate transactions offering the same custom token for the 
	block. Thus, the value $v_i$ of a candidate 
	transaction is expressed as a function of what has already been selected 
	for the block, $v_i(B_{i-1}): Candidate Block \rightarrow 
	\mathit{Value}$, where $B_{i-1} 
	\subseteq \{1,2,...,i-1\}$ and $v_i(\emptyset) = v_{io}$ is the initial 
	value of the offer and $0 \leq v_i(B_{i-1}) \leq v_{oi}$. Finally, the 
	$\mathtt{utility}$ function that we want to maximize is defined as $utility 
	= 
	\sum_{i \in B_n}^{}v_i(B_{i-1})$, where $B_{i-1}$ is the solution to the 
	$TxSelection(i-1,S_B - 
	\sum_{j=i}^{n}s_j,M-\{i,...,n\},R-\sum_{j=i}^{n}l_j)$ problem.
	
\end{definition}

\subsection{Dynamic programming}

We start with the presentation of an an optimal solution to 
the 
transaction selection problem. It is a variation of the dynamic programming 
solution to the 0-1 knapsack problem~\cite{algorithms}. It is important to note 
that we want conventional transactions and transaction batch offers to be 
comparable \emph{only} with respect to the value offered and their size. We do not want to view liability as another constraint to 
the knapsack problem, because this would favor zero liability candidate 
transactions (i.e., conventional transactions) over Babel offers. The liability 
aspect of the offer has already been considered in the value calculation of
the conversion function from a $\mathtt{BabelOffer}$ to a 
$\mathtt{CandidateTransaction}$, as shown in the indicative conversion 
formula above. 

\subsection{Optimal algorithm of the transaction selection problem}
\label{sec:alg-optimal}

The optimal algorithm presented in Algorithm~\ref{alg-tx-selection} proceeds as follows. Initially, we order the 
candidate transactions of $M$ in descending order of their (initial) value per 
size ratio 
$v_{io}/s_i, i = 1,2,...n$. We maintain an array $U[i], i = 1,2,...n$. Each 
entry $U[i]$ is a list of tuples of the form $(t_s,t_v,r,b)$. A tuple 
$(t_s,t_v,r,b)$ 
in the list $U[i]$ indicates that there is a block $B$ assembled from the first 
$i$ 
candidate transactions that uses space exactly $t_s \leq S_B$, has a total 
value 
exactly $utility(B) = t_v \leq \sum_{i=1}^{n}v_{oi}$, has a residual amount of 
primary currency to be spent on liabilities exactly $r \leq R$ and has a 
\emph{participation bit} $b$ indicating if transaction $i$ is included in $B$, 
or not.

This list does not contain all possible such tuples, but instead keeps track of 
only the most efficient ones. To do this, we
introduce the notion of one tuple \emph{dominating} another one; a tuple 
$(t_s,t_v,r,b)$ dominates another tuple $(t_s',t_v',r',b')$, if $t_s\leq t_s'$ 
and $t_v\geq t_v'$; that is, the solution indicated by the tuple 
$(t_s,t_v,r,b)$ uses no more space
than $(t_s',t_v',r',b')$, but has at least as much value. Note that domination 
is a 
transitive property;
that is, if $(t_s,t_v,r,b)$ dominates $(t_s',t_v',r',b')$ and 
$(t_s',t_v',r',b')$ dominates 
$(t_s'',t_v'',r'',b'')$, then $(t_s,t_v,r,b)$ also dominates
$(t_s'',t_v'',r'',b'')$. We will ensure that in any list, no tuple dominates 
another one; this means that we
can assume each list $U[i]$ is of the form 
$[(t_{s1},t_{v1},r_1,b_1),...,(t_{sk},t_{vk},r_k,b_k)]$ with $t_{s1} < t_{s2} < 
... < t_{sk}$ and 
$t_{v1} < t_{v2} < ... < t_{vk}$. Since every list $U[i], i = 1,2,...,n$ does 
not 
include dominating tuples and also the sizes of the transactions are integers 
and 
so are their values, then we can see that the maximum length of such a list 
is $min(S_B+1, V_o + 1)$, where $V_o = 
\sum_{i=1}^{n}v_{oi}$.

Algorithm~\ref{alg-tx-selection} starts out with the initialization of list 
$U[1]$ (line 2) and then iterates through all $n-1$ transactions (lines 3-10). 
In 
each iteration $j$, we initially set $U[j] \longleftarrow U[j-1]$ after turning 
off the participation bit in all tuples (lines 4-5). Then for each tuple 
$(t_s,t_v,r,b) \in U[j-1]$, we also add the tuple 
$(t_s+s_j,t_v+v_j(B_{j-1}),r-l_j, 1)$ 
to the list, if $t_s + s_j \leq S_B \land r - l_j \geq R$; that is, if by 
adding 
transaction $j$ to the corresponding subset, we do not surpass the total 
available size $S_B$ and do not deplete our reserve $R$ for liabilities (lines 
6-9). Note that the value of transaction $j$ at this point is determined by 
the contents of the corresponding block $B_{j-1}$ through the function call 
$v_j(B_{j-1})$. To this end, in lines 14-22 we provide a function that returns 
the block corresponding to a specific tuple. We finally remove from $U[j]$ 
all dominated tuples by
sorting the list with respect to their space component, retaining the best 
value for each space
total possible, and removing any larger space total that does not have a 
corresponding larger value (line 10). We return the maximum total value from 
the list $U[n]$ along with the corresponding block $B_n$ (lines 11-13).
\begin{algorithm}
	\scriptsize

	\KwIn{A set $M$ of candidate transactions $M = \{tx_1, tx_2, ...,tx_n\}$,
		where $tx_i = (i, v_i(B_{i-1}), l_i, s_i)$ for $i = 1,...,n$ according
		to
		definition \ref{def:candidatetx}}
	\KwIn{An amount of primary currency available for covering liabilities,
		called
		the	reserve	$R$.}
	\KwIn{An available block size $S_B$}
	\KwIn {A utility function
		$util = \sum_{i \in B_n}^{}v_i(B_{i-1})$}

	\KwOut{$(B, util(B), res)$: A candidate block $B \subseteq M$ such that
		$util(B) > util(B') \forall B' \subseteq M$, the value of this block
		($util(B)$) and a residual amount
		$res$ from the reserve $R$ such that $res \geq 0$}

	\tcc{Assume array U[i]: Array[List[(Size, Value, Liability, Bit)]], $i =
		1,...n$}

	order transactions in $M$ in descending order of $v_{io}/s_i$, $i =
	1,2...,n$

	$U[1] \longleftarrow [(0,0,R,0), (s_1,v_{1o},R-l_1, 1)]$

	\For{$j = 2$ to $n$}{
		$baseList \longleftarrow$ copy list $U[j-1]$ with zero
		participation bits for all tuples

		$U[j] \leftarrow $ baseList

		\ForEach{$(t_s,t_v,r,b) \in \mathtt{baseList}$}{
			\If{$t_s + s_j \leq S_B \land r - l_j \geq R$}{
				$B_{j-1} \longleftarrow getBlock(U,j-1, t_s)$

				Add tuple $(t_s+s_j,t_v+v_j(B_{j-1}),r-l_j, 1)$ to $U[j]$

			}
		}
		Remove dominating pairs from list $U[j]$
	}
	$(S_{final}, V_{max}, residual, b) \longleftarrow max_{(s,v,r) \in U[n]}(v)$

	$B_n \longleftarrow getBlock(U,n,S_{final})$

	return ($B_n$, $V_{max}$, residual)

	\tcp{-----------------------------------------------------------------}

	getBlock(U: Array[List[(Size, Value, Liability, Bit)]], n:$Tx_{id}$,
	$t_{sn}$:
	Size)  return CandidateBlock

	$B \longleftarrow []$

	$t_s \longleftarrow t_{s_n}$

	\For{i = $n$ down to 1}{
		$(t_{si}, t_{vi}, r_i, b_i) \longleftarrow getTuple (U[i], t_s)$

		\If{$b_i == 1$}{
			$B \longleftarrow i : B$\ \ \tcp{":" is list construction}

			$t_s \longleftarrow t_s - t_{si}$
		}
	}
	return $B$
	\caption{Transaction selection algorithm for a block (Optimal Solution).}
	\label{alg-tx-selection}
\end{algorithm}

\begin{theorem}\label{th:optimality}
	Algorithm \ref{alg-tx-selection} correctly computes the optimal value for 
	the transaction selection problem.
\end{theorem}
The proof of the theorem is contained in \longelse{Appendix \ref{sec:proof-of-optimality}}{the unabridged paper~\cite[Appendix~E]{longversion}}.

\subsection{Polynomial approximation}

Since we iterate through all available $n$ transactions and in each iteration 
we process a list of length $min(S_B+1, V_o + 1)$, where $V_o = 
\sum_{i=1}^{n}v_{oi}$, we can see that algorithm \ref{alg-tx-selection} 
takes $O(n\ min(S_B, V_o))$ time. This is not a polynomial-time algorithm, since
we assume that all input numbers are encoded in binary; thus, the size of the 
input number $S_B$ is essentially $log_2 S_B$, and so the running time 
$O(nS_B)$ is  exponential in the size of the input number $S_B$, not 
polynomial. Based on the intuition that if the maximum value $V_o$ was bounded 
by a polynomial in $n$, the running time will indeed be a polynomial in 
the input size, we now propose an approximation algorithm for the 
transaction selection problem that runs in polynomial time and is based on 
a well-known fully polynomial approximation scheme of the 0-1 knapsack problem~\cite{knapsack}.

The basic intuition of the approximation algorithm is that if we round the 
(integer) values of the candidate transactions to $v_i'(B_{i-1}) = \lfloor 
v_i(B_{i-1})/\mu$$\rfloor$, where $0 \leq v_i'(B_{i-1}) \leq  \lfloor 
v_{io}/\mu\rfloor = v_{io}'$ and run Algorithm~\ref{alg-tx-selection} with 
values $v_i'$ instead of $v_i$, then by an appropriate 
selection of $\mu$, we could 
bound the maximum value $V_o' = \sum_{i=1}^{n}v_{io}'$ by a polynomial in $n$ 
and return a solution that 
is  at least $(1 - \epsilon)$ times the value of the optimal solution (OPT). In 
particular, if we choose $\mu = \epsilon v_{omax} / n$, where $v_{omax}$ is the 
maximum value of a transaction; that is, $v_{omax} = max_{i \in M} (v_{oi})$. 
Then, for the total maximum value $V_o'$, we have $V_o' = \sum_{i=1}^{n}v_{io}' 
= \sum_{i=1}^{n}\lfloor \frac{v_{io}}{\epsilon v_{omax}/n}\rfloor = 
O(n^2/\epsilon)$.  Thus, the running time of the algorithm is $O(n\ min(S_B, 
V_o')) = O(n^3 /\epsilon)$ and is bounded by a polynomial in $1/\epsilon$. Algorithm~\ref{alg-tx-selection_apprx} comtains our 
approximate algorithm 
for the transaction selection problem. Essentially, we run algorithm 
\ref{alg-tx-selection} for the problem instance 
$\mathtt{TxSelection}(n, S_B, M', R)$, where $M' = \{tx_1', tx_2', 
...,tx_n'\}$, 
and $tx_i' = (i, v_i'(B_{i-1}), l_i, s_i)$ for $i = 1,...,n$ We can now prove 
that this algorithm 
returns a solution whose value is at least $(1-\epsilon)$
times the value of the optimal solution.
\begin{algorithm}
	\scriptsize

	\KwIn{A set $M$ of candidate transactions $M = \{tx_1, tx_2, ...,tx_n\}$,
		where $tx_i = (i, v_i(B_{i-1}), l_i, s_i)$ for $i = 1,...,n$ according
		to
		definition \ref{def:candidatetx}}
	\KwIn{An amount of primary currency available for covering liabilities,
		called
		the	reserve	$R$.}
	\KwIn{An available block size $S_B$}
	\KwIn {A utility function
		$util = \sum_{i \in B_n}^{}v_i(B_{i-1})$}
	\KwIn{The acceptable error $\epsilon$ from the optimal solution, where $0 <
		\epsilon < 1$}

	\KwOut{$(B, util(B), res)$: A candidate block $B$ such that $util(B) >
		util(B') \forall B' \subseteq M$, the value of this block
		($util(B)$) and a residual amount
		$res$ from the reserve $R$ such that $res \geq 0$}

	\caption{Transaction selection algorithm for a block (Approximate
		Solution).}
	\label{alg-tx-selection_apprx}

	$v_omax \longleftarrow max_{i \in M} (v_{oi})$

	$\mu \longleftarrow \epsilon v_{omax} / n$

	$v_i'(B_{i-1}) \longleftarrow \lfloor v_i(B_{i-1})/\mu$$\rfloor$ for
	$i=1,2,...,n$

	run algorithm \ref{alg-tx-selection} for the problem instance
	$\mathtt{TxSelection}(n, S_B, M', R)$, where $M' = \{tx_1', tx_2',
	...,tx_n'\}$,
	and $tx_i' = (i, v_i'(B_{i-1}), l_i, s_i)$ for $i = 1,...,n$
\end{algorithm}

\begin{theorem} \label{th:approx-solution}
	Algorithm \ref{alg-tx-selection_apprx} provides a solution which is at 
	least $(1-\epsilon)$ times the value of OPT.
\end{theorem}
The proof of the theorem is in \longelse{Appendix~\ref{sec:proof-of-approx-solution}}{the unabridged paper~\cite[Appendix~F]{longversion}}.

%
\section{Related Work}
\label{sec:related-work}

Babel fees are enabled by swap outputs based on limited-lifetime liabilities. These swaps, once being proposed (as part of a complete transaction), can be resolved unilaterally by the second party accepting the swap as elaborated in \longelse{Appendix~\ref{sec:swaps}}{the unabridged paper~\cite[Appendix~B]{longversion}}.

\subsubsection{Atomic swaps and collateralized loans.}
Atomic swaps (which may be used to pay for fees) often go via an exchange, including for Ethereum ERC-20 tokens~\cite{kyber} and Waves' custom natives~\cite{wavesdex}, as well as multi-blockchain exchanges based on atomic swaps~\cite{idex,atomicdex}. These exchanges come in varying degrees of decentralisation. Atomic swaps are also
used for swapping or auctioning assets across chains~\cite{herlihy2018atomic,summa}. Our proposal is fully decentralized and single-chain. It allows transactions carrying swap or fee-coverage offers to be disseminated directly via the
blockchain network (because they are fully-formed transactions), without any off-chain communication.

A notable difference between our swap mechanism and some layer-2 DEX solutions, such as Ethereum's Uniswap~\cite{uniswap} and SwapDEX~\cite{swapdex},
is that these require proof of liquidity (ie. assets locked in a contract), as well as contract-fixed exchange
rates. Our proposal enables users to accept the optimal number of exchange
offers without an obligation to have liquidity or to accept them. Users are also free to
choose and change their exchange rates at any time, without on-chain actions.

Our limited-lifetime liabilities are a sort of loan, but one that is resolved before it is even recorded on the ledger. There is also work on ledger-based loans~\cite{black2019atomic,maker}, but this leads to rather different challenges and mechanisms. In particular,
the liabilities we propose do not require collateral backing (as they are resolved within a single batch). Moreover,
unlike either atomic swaps or collateralized loans, our mechanism requires no actions from the user after submitting
a swap offer or fee-less transaction to the network.

These mechanisms, while having some capacity to address some of the same shortcomings as
the babel fees mechanism, are usually a combination of off-chain solutions and layer-2 (via smart contracts) are quite different
from the single-chain, ledger-integrated proposal we provided.

\subsubsection{Child pays for parent.}
The UTXO model enforces a partial ordering on transactions that can be taken advantage of
to encourage block producers to include less desirable (smaller-fee) transactions
in a block by also disseminating a higher-fee transaction that depends on the undesirable
transation. This is known as a child-pays-for-parent technique~\cite{cpfp}. Like ours,
it deals with fully formed transactions, and requires no further input from the author of
the small-fee transaction. The solution
we propose, however, is geared towards a ledger model where transaction validation rules enforce a minimum
fee, so any transaction that does not pay it (via liabilities or directly) will be rejected
regardless of whether a high-fee transaction depends on it.

\subsubsection{Ethereum.}
Ethereum's Gas Station Network (GSN)~\cite{gsn} infrastructure
consists of (a) a network of nodes listening for meta-transactions (trans\-ac\-tion-like
  requests to cover transaction fees), which turn these requests into complete
  transactions, with fees covered by the relay node, and (b) an interface that contracts must implement in order for the relay
  nodes to use this contract's funds to subsidize the transaction fees.

Babel fees are simpler as they don't require the following
(all of which the GSN relies on): (1) disseminating of partially formed (meta-)transactions on a separate network, (2) adding infrastructure, such as relays, relay hubs, and a separate communication network, (3) any changes to smart contracts to allow them to participate, (4) submitting transactions to make or update fee-covering or exchange offers, (5) any further action from the user after submitting a transaction that
  requires its fees to be covered, and (6) pre-paying for the fee amounts contracts are able to cover.


Another solution for processing transactions without any primary currency
included to cover fees, called Etherless Ethereum Tokens, is propsed in \cite{eet}.
This approach includes a formal composability framework (including formal proofs
of important properties), requires notably less gas consumption, and offers a much more
seamless user experience than the GSN. However, it still relies on the off-chain
dissemination of meta-transactions, and
requires changes to smart contracts to opt in to participation, as well as fix an exchange rate. 

\subsubsection{Algorand.}
Algorand is an account-based cryptocurrency which supports custom native tokens. It
provides users with a way to perform atomic transfers (see \cite{algorand}).
An atomic transfer requires combining unsigned transactions into a single group transaction,
which must then be signed by each of the participants of each of the transactions included.
This design allows users to perform, in particular, atomic swaps, which might be used to
pay fees in non-primary currencies.

As with our design, the transactions get included into the ledger in batches.
Unlike Babel fees, however, incomplete transactions cannot be sent off
to be included in the ledger without any further involvement of the transaction author.

\subsubsection{Debt representation in UTXO blockchains.}
There are similarities between the debt representation proposal presented
in~\cite{utxo-debt} and the mechanism we propose, the main one being the
idea of representing debt as special inputs on an \UTXO\ ledger. Unlike
the debt model we propose, the model presented in that paper allows debt to be recorded in a persistent
way on the ledger.  As we prevent liabilities to ever enter the ledger state, we side step the main issues discussed in~\cite{utxo-debt}, including the need for managing permissions
for issuing debt on the ledger, and therefore also for the trust users may be
obligated to place in the debt issuer, and vice-versa. The possibility of
unresolved debt remaining on the ledger (and therefore inflation) is a concern that needs to be taken seriously in this case.

Debt recorded on the ledger state (and outside a transaction batch) enables functionality that we cannot support with limited liabilities. Moreover, if a debt-creating transaction
is complete and ready to be applied to the ledger, all nodes are able to explicitly
determine the validity of this transaction. This way, these transaction can be relayed
by the existing network,
without any special consideration for their potential to be included in a batch,
and by who.

Another key difference between the two proposals is that ours assumes an underlying
multi-asset ledger, so that the debt-outputs have another major interpretation ---
they also serve as offers for custom token fee coverage, as well as swaps. Finally, the ledger
we propose treats debt outputs and inputs in a uniform way, rather than
in terms of special debt transactions and debt pools, which result in potentially complicated special cases.

\subsubsection{Stellar DEX.}
The Stellar system~\cite{stellar}
supports a native, ledger-implemented DEX to provide swap functionality (and therefore, custom token fee payment).

In the Stellar DEX, offers posted by users are stored on the ledger.
A transaction may attempt an exchange of any asset for any other asset, and
will fail if this exchange is not offered. This approach requires submitting transactions
to manage a user's on-chain offers, and also requires all exchanges to be exact --- which means no
overpaying is possible to get one's bid selected. A transaction may attempt to
exchange assets that are not explicitly listed as offers in exchange for each other on the DEX.
The DEX, in this case, is searched for a multi-step path to exchanging these assets via intermediate offers.
This is not easily doable using the approach we have presented.

A DEX of this nature is susceptible to front-running. In our case, block issuers
are given a permanent advantage in resolving liability transactions over non-block-issuing
users. Among them, however, exactly one may issue the next block, including the
liabilities they resolved.


\bibliographystyle{splncs04}
\bibliography{liabilities}

\begin{thebibliography}{10}
\providecommand{\url}[1]{\texttt{#1}}
\providecommand{\urlprefix}{URL }
\providecommand{\doi}[1]{https://doi.org/#1}

\bibitem{gsn}
{ Yoav Weiss and Dror Tirosh and Alex Forshtat}: {EIP-1613: Gas stations
  network}. \url{https://eips.ethereum.org/EIPS/eip-1613} (2018)

\bibitem{algorand}
{Algorand Team}: {Algorand Developer Documentation}.
  \url{https://developer.algorand.org/docs/} (2021)

\bibitem{eet}
Andrews, J., Ciampi, M., Zikas, V.: Etherless ethereum tokens: Simulating
  native tokens in ethereum. Cryptology ePrint Archive, Report 2021/766 (2021),
  \url{https://ia.cr/2021/766}

\bibitem{black2019atomic}
Black, M., Liu, T., Cai, T.: Atomic loans: Cryptocurrency debt instruments
  (2019)

\bibitem{eutxoma}
Chakravarty, M.M.T., Chapman, J., MacKenzie, K., Melkonian, O., M{\"{u}}ller,
  J., Jones, M.P., Vinogradova, P., Wadler, P.: Native custom tokens in the
  extended {UTXO} model. In: Leveraging Applications of Formal Methods,
  Verification and Validation: Applications - 9th International Symposium on
  Leveraging Applications of Formal Methods, ISoLA 2020, Rhodes, Greece,
  October 20-30, 2020, Proceedings, Part {III}. LNCS, vol. 12478 (2020)

\bibitem{utxoma}
Chakravarty, M.M.T., Chapman, J., MacKenzie, K., Melkonian, O., M{\"{u}}ller,
  J., Jones, M.P., Vinogradova, P., Wadler, P., Zahnentferner, J.:
  {UTXO}\(_{\mbox{ma}}\): {UTXO} with multi-asset support. In: Leveraging
  Applications of Formal Methods, Verification and Validation: Applications -
  9th International Symposium on Leveraging Applications of Formal Methods,
  ISoLA 2020, Rhodes, Greece, October 20-30, 2020, Proceedings, Part {III}.
  LNCS, vol. 12478 (2020)

\bibitem{eutxo}
Chakravarty, M.M.T., Chapman, J., MacKenzie, K., Melkonian, O., {Peyton Jones},
  M., Wadler, P.: The {Extended UTXO} model. In: Proceedings of Trusted Smart
  Contracts (WTSC). LNCS, vol. 12063. Springer (2020)

\bibitem{utxo-debt}
Chiu, M., Kalabić, U.: Debt representation in {UTXO} blockchains. In:
  Financial Cryptography and Data Security 2021 (2021)

\bibitem{algorithms}
Cormen, T.H., Leiserson, C.E., Rivest, R.L., Stein, C.: Introduction to
  Algorithms, Third Edition. The MIT Press, 3rd edn. (2009)

\bibitem{backbone}
Garay, J.A., Kiayias, A., Leonardos, N.: The bitcoin backbone protocol:
  Analysis and applications. In: Oswald, E., Fischlin, M. (eds.) Advances in
  Cryptology - {EUROCRYPT} 2015 - 34th Annual International Conference on the
  Theory and Applications of Cryptographic Techniques, Sofia, Bulgaria, April
  26-30, 2015, Proceedings, Part {II}. LNCS, vol.~9057, pp. 281--310. Springer
  (2015)

\bibitem{herlihy2018atomic}
Herlihy, M.: Atomic cross-chain swaps (2018)

\bibitem{idex}
{IDEX Team}: {IDEX documentation}. \url{https://docs.idex.io/} (2021)

\bibitem{knapsack}
Kellerer, H., Pferschy, U., Pisinger, D.: Knapsack Problems. Springer Berlin
  Heidelberg, Berlin, Heidelberg (2004)

\bibitem{atomicdex}
{Komodo Team}: {AtomicDEX documentation}.
  \url{https://developers.komodoplatform.com/basic-docs/atomicdex/} (2021)

\bibitem{kyber}
{Kyber Team}: {Kyber: An On-Chain Liquidity Protocol}.
  \url{https://files.kyber.network/Kyber_Protocol_22_April_v0.1.pdf} (2019)

\bibitem{summa}
Prestwich, J.: Cross-chain auctions via bitcoin double spends.
  \url{https://medium.com/summa-technology/summa-auction-bitcoin-technical-7344096498f2}
  (2018)

\bibitem{cpfp}
Project, B.: Developer guides. \url{https://developer.bitcoin.org/devguide/}
  (2020)

\bibitem{stellar}
{Stellar Development Foundation}: {Stellar Development Guides}.
  \url{https://developers.stellar.org/docs/} (2020)

\bibitem{swapdex}
{SWAPDEX Team}: {SWAPDEX : Decentralized Finance is the Future}.
  \url{https://swapdex.net/whitepaper/SWAPDEX.pdf} (2020)

\bibitem{maker}
Team, M.: The maker protocol: Makerdao's multi-collateral dai (mcd) system.
  \url{https://makerdao.com/en/whitepaper/} (Retrieved May 26, 2021)

\bibitem{uniswap}
Team, U.: Uniswap whitepaper.
  \url{https://hackmd.io/@HaydenAdams/HJ9jLsfTz#%F0%9F%A6%84-Uniswap-Whitepaper}
  (2019)

\bibitem{wavesdex}
{Waves.Exchange Team}: {Waves.Exchange Documentation}.
  \url{https://docs.waves.exchange/en/waves-exchange/} (2021)

\bibitem{zahnentferner:chimeric-ledgers}
Zahnentferner, J.: Chimeric ledgers: Translating and unifying {UTxO}-based and
  account-based cryptocurrencies. {IACR} Cryptology ePrint Archive
  \textbf{2018}, ~262 (2018), \url{http://eprint.iacr.org/2018/262}

\end{thebibliography}

\iflong
\clearpage
\appendix

\section{Definitions supporting the formal ledger rules}
\label{sec:utxoma}

The \UTXOll\ ledger rules presented in Section~\ref{sec:ledger-rules} are based on those of \UTXOma~\cite{utxoma}. This appendix summarises supporting definitions for ease of reference. There is, however, one notable simplification in the system that we use here compared to the original \UTXOma\ system. The original \UTXOma\ system, while not supporting general-purpose scripts for smart contracts, does support a special-purpose language for defining asset policy script as well as multisig and timed \UTXO\ validator scripts. In the present work, we keep the asset policy script, but we restrict \UTXO\ outputs to be simply pay-to-pubkey outputs. This is to keep the presentation simpler and because the additional functionality of \UTXOma\ doesn't lead to additional insight in the context of the present paper.

\begin{figure}[h]
  \begin{displaymath}
    \begin{array}{rll}
      \B{} && \mbox{the type of Booleans}\\
      \N{} && \mbox{the type of natural numbers}\\
      \Z{} && \mbox{the type of integers}\\
      \H{} && \mbox{the type of bytestrings: } \bigcup_{n=0}^{\infty}\{0,1\}^{8n}\\
      (\phi_1 : T_1, \ldots, \phi_n : T_n) && \mbox{record type with fields $\phi_i$ of types $T_i$}\\
      t.\phi && \mbox{the value of $\phi$ for $t$,}\\
             && \mbox{where $t$ has type $T$ and $\phi$ is a field of $T$}\\
      \Set{T} && \mbox{the type of (finite) sets over $T$}\\
      \List{T} && \mbox{the type of lists over $T$, }\\
               && \mbox{with $\_[\_]$ as indexing and $|\_|$ as length}\\
      h::t && \mbox{the list with head $h$ and tail $t$}\\
      h++t && \mbox{list concatenation}\\
      x \mapsto f(x) && \mbox{an anonymous function}\\
      \hash{c} && \mbox{cryptographic collision-resistant hash of $c$}\\
      \Interval{A} && \mbox{type of intervals over totally-ordered set $A$}\\
      \FinSup{K}{M} && \mbox{type of finitely supported functions}\\
                    && \mbox{from a type $K$ to a monoid $M$}
    \end{array}
  \end{displaymath}
  \caption{Basic types and notation}
  \label{fig:basic-types}
\end{figure}
Figure~\ref{fig:basic-types} includes some basic types and notation.
Crucial are finitely-supported functions, which provide the algebraic structure underpinning token bundles on a multi-asset ledger.

\paragraph{Finitely-supported functions.}
\label{sec:fsfs}
We model token bundles as finitely-supported functions.
If $K$ is any type and $M$ is a monoid with identity element $0$, then a function $f: K \rightarrow M$ is \textit{finitely supported} if $f(k) \ne 0$ for only finitely many $k \in K$.
More precisely, for $f: K \rightarrow M$ we define the \textit{support} of $f$ to be
$\supp(f) = \{k \in K : f(k) \ne 0\}$
and
$\FinSup{K}{M} = \{f : K \rightarrow M : \left|\supp(f)\right| < \infty \}$.

If $(M,+,0)$ is a monoid then $\FinSup{K}{M}$ also becomes a monoid if we define addition pointwise (i.e., $(f+g)(k) = f(k) + g(k)$), with the identity element being the zero map.
Furthermore, if $M$ is an abelian group then $\FinSup{K}{M}$ is also an abelian group under this construction, with $(-f)(k) = -f(k)$.
Similarly, if $M$ is partially ordered, then so is $\FinSup{K}{M}$ with comparison defined pointwise: $f \leq g$ if and only if $f(k) \leq g(k)$ for all $k \in K$.

It follows that if $M$ is a (partially ordered) monoid or abelian group then so is $\FinSup{K}{\FinSup{L}{M}}$ for any two sets of keys $K$ and $L$.
We will make use of this fact in the validation rules presented later in the paper (see Figure~\ref{fig:validity}).
Finitely-supported functions are easily implemented as finite maps, with a failed map lookup corresponding to returning 0.

\subsection{Ledger types}

\begin{figure*}
  \begin{displaymath}
    \begin{array}{rll}
      \multicolumn{3}{l}{\textsc{Ledger primitives}}\\
      \Quantity && \mbox{an amount of currency, forming an abelian group (typically \Z{})}\\
      \AssetID && \mbox{a type consisting of identifiers for individual asset classes}\\
      \Tick && \mbox{a tick}\\
      \Address && \mbox{an ``address'' in the blockchain}\\
      \TxId && \mbox{the identifier of a transaction}\\
      \txId : \utxotx \rightarrow \TxId && \mbox{a function computing the identifier of a transaction}\\
      \lookupTx : \Ledger \times \TxId \rightarrow \utxotx && \mbox{retrieve the unique transaction with a given identifier}\\
      \verify : \PublicKey\times\H\times\H \rightarrow \B && \mbox{signature verification}\\
      \keyAddr : \PublicKey \rightarrow \Address && \mbox{the address of a public key}\\
      \Script && \mbox{forging policy scripts}\\
      \scriptAddr : \Script \rightarrow \Address && \mbox{the address of a script}\\
      \applyScript{\_} : \Script \to (\Address \times \utxotx \times \Set{\Output}) \to
      \B && \mbox{apply script inside brackets to its arguments}\\
    \\
    \multicolumn{3}{l}{\textsc{Ledger types}} \\
    \PolicyID &=& \Address \qquad \mbox{(an identifier for a custom asset)}\\
    \Signature &=& \H\\
    \\
    \AssetID   &=& (pid: \PolicyID, aName: \AssetName)\\
    \\
    \Quantities   &=& \FinSup{\AssetID}{\Quantity}\\
    \\
    \Output &=& (\addr: \Address, \val: \Quantities)\\
    \\
    \OutputRef &=& (\txrefid: \TxId, \idx: \s{Int})\\
    \\
    \Input &=& ( \outputref : \OutputRef\\
             & &\ \key: \PublicKey)\\
    \\
    \utxotx &=&(\inputs: \Set{\Input},\\
               & &\ \outputs: \List{\Output},\\
               & &\ \validityInterval: \Interval{\Tick},\\
               & &\ \forge: \Quantities\\
               & &\ \scripts: \Set{\Script},\\
               & &\ \sigs: \Set{\Signature})\\
    \\
    \Ledger &=&\!\List{\utxotx}\\
    \end{array}
  \end{displaymath}
  \caption{Ledger primitives and basic types}
  \label{fig:ledger-types}
\end{figure*}
Figure~\ref{fig:ledger-types} defines the ledger primitives and types that we need to define the \UTXOll\ model.
All outputs use a pay-to-pubkey-hash scheme, where an output is locked with the hash of key of the owner. We use a simple scripting language for forging policies, which we don't detail here any further --- please see~\cite{utxoma} for details.
We assume that each transaction has a unique identifier derived from its value by a hash function. This is the basis of the $\lookupTx$ function to look up a transaction, given its unique identifier.

\paragraph{Token bundles.}

We generalise per-output transferred quantities from a plain \Quantity\ to a bundle of \Quantities.
A \Quantities{} represents a token bundle: it is a mapping from a policy and an \emph{asset}, which defines the asset class, to a \Quantity{} of that asset.\footnote{
  We have chosen to represent \Quantities{} as a finitely-supported function whose values are themselves finitely-supported functions (in an implementation, this would be a nested map).
  We did this to make the definition of the rules simpler (in particular Rule~\ref{rule:forging}).
  However, it could equally well be defined as a finitely-supported function from tuples of \PolicyID{}s and \AssetID{}s to \Quantity{}s.
}
Since a \Quantities\ is indexed in this way, it can represent any combination of tokens from any assets (hence why we call it a token \emph{bundle}).

\paragraph{Asset groups and forging policy scripts.}

A key concept is the \emph{asset group}.
An asset group is identified by the hash of special script that controls the creation and destruction of asset tokens of that asset group.
We call this script the \emph{forging policy script}.

\paragraph{Forging.}

Each transaction gets a $\forge$ field, which simply modifies the required balance of the transaction by the $\Quantities$ inside it: thus a positive $\forge$ field indicates the creation of new tokens.
$\Quantities$ in forge fields can
also be negative, which effectively burns existing tokens.

Additionally, transactions get a $\scripts$ field holding a set of forging policy scripts: \(\Set{\Script}\).
This provides the forging policy scripts that are required as part of validation when tokens are minted or destroyed (see Rule~\ref{rule:forging} in Figure~\ref{fig:conditional-validity}). The forging scripts of the assets being forged are
executed and the transaction is only considered valid if the execution of the script returns $\true$.
A forging policy script is executed in a context that provides access to the main components of the forging transaction, the UTXOs it spends, and the policy ID.
The passing of the context provides a crucial piece of the puzzle regarding self-identification: it includes the script's own $\Policy$, which avoids the problem of trying to include the hash of a script inside itself.

\paragraph{Validity intervals.}
\label{para:validity-intervals}

A transaction's \emph{validity interval} field contains an interval of ticks (monotonically increasing units of time, from~\cite{eutxo}).
The validity interval states that the transaction must only be validated if the current tick is within the interval. The validity interval, rather than the actual current chain tick value, must be used
for script validation. In an otherwise valid transaction, passing the current
tick to the evaluator
could result in different script validation outcomes at different ticks, which
would be problematic.

\section{Other applications of limited liabilities} \label{sec:appx:other-applications}

\subsubsection{Swaps.}
\label{sec:swaps}
As discussed, we use liabilities in babel fees to form transaction outputs that 
represent atomic swaps --- we call those \emph{swap outputs}. We do this by 
including a liability (negative token value) together with an asset (positive 
token value). Whoever consumes such an output effectively swaps the tokens 
described by the liability for those constituting the asset.

\UTXO\ ledgers do already support atomic swaps natively by way of transactions 
consuming two inputs carrying two different assets and swapping the keys under 
which these two assets are locked. Such a \emph{swap transaction} does require 
both parties to sign that  transaction (to authorise the consumption of each 
parties share in the swap). Moreover, both parties to the swap must be known at 
swap transaction creation time.

Liabilities enable us to break this monolithic, cooperative process into a 
non-interactive two-stage process. The first party creates a liability 
transaction consuming only it's own share of the swap and bundling it with a 
liability of the expected return in a swap output. The second party combines 
the liability transaction with a second transaction that resolves the liability 
and completes the swap. The second party can do that solely on the basis of the 
first transaction without any additional need to interact with the first party. 

Our batch example from Section~\ref{sec:tx_batch} demonstrates this pattern. 
The first output of $t_1$, namely 
\[
(\addr: \emptyset, \val: \{ T_1 \mapsto -5, T_2 \mapsto 10 \}),
\]
is the swap output, offering to swap $10 T_2$ for $5 T_1$.

\subsubsection{Service payments.}
Extending the concept of swaps from exchanging assets to exchanging assets for 
information. In the Extended \UTXO\ model~\cite{eutxoma}, which facilitates 
complex smart contracts on a \UTXO\ ledger, transaction outputs also include a 
data component. This can, for example, be used to communicate information from 
an off-chain oracle. Liabilities included with such an output can serve as 
payment for consuming such an output with the data.

\subsubsection{Indivisiblity.}
Transaction batches are different to signed transaction groups proposed for 
some ledgers, such as, for example, Algorand~\cite{algorand}. To create a 
signed transaction group, all component transactions need to be known and the 
group signed as a whole. If multiple component transactions are created by 
different parties, these parties need to cooperate to create the group 
transaction. The benefit of such a signed group is that it is indivisible.

The transaction batches that we propose are different. They are not inherently 
indivisible. For example, a batch comprising two transactions $t_1$ and $t_2$, 
where the latter consumes an output of the former, may be included in the 
ledger as a whole, but unless there is a liability involved in the output of 
$t_1$ consumed by $t_2$, we could also split the batch and simply submit $t_1$ 
on its own.

And even if $t_1$'s output in includes a liability, while this prevents $t_1$ 
to go onto the ledger by itself, it still leaves the possibility of replacing 
$t_2$ by another transaction that resolves $t_1$'s liability. In other words, 
while a liability may prevent a prefix of a batch to be fully valid on its own, 
it may not prevent swapping out a suffix of transactions.

Nevertheless, we can use spending conditions on outputs with liabilities to 
exert control over liability-resolving transactions in a batch. With 
pay-to-pubkey outputs, we can control who may create these transactions and 
within the Extended \UTXO\ model~\cite{eutxoma}, we can use script code to 
exert fine-grained control over these transactions.

\section{Account-based ledgers} \label{sec:appx:account-based-ledgers}

In this paper, we explain limited liabilities in the context of a \UTXO\ ledger 
model, building on the native custom token extension for \UTXO, called \UTXOma\ 
introduced in~\cite{utxoma}. We like to emphasize, though, that both the native 
custom token extension of \UTXOma\ and the concept of limited liabilities from 
the present paper can equally well be applied to an account-based ledger --- 
this might not come as a surprise, given that one accounting model can be 
translated into the other~\cite{zahnentferner:chimeric-ledgers}.

\subsubsection{Native custom tokens.}
The core ideas of native custom tokens in \UTXOma\ are (a) the generalisation 
of integral values in \UTXO{}s to token bundles and (b) the use of 
policy-controlled forge fields. In the context of an account-based ledger, 
Point (a) translates to accounts that hold an entire bundle of tokens instead 
of just coins of a single currency. Just like in the \UTXOma\ model, these 
token bundles can be represented as finitely-supported functions using the same 
group structure as \UTXOma\ for value calculations. In a similar manner, 
transactions transferring value from one account to another now transfer entire 
token bundles, which get deducted from the source account and added to the 
target account, with the constraint that none of the bundle components of the 
source account may become negative as a result.

To realise Point (b), we extent transfer transactions with the same sort of 
forge field as in \UTXOma. This includes the same conditions on the use of 
asset policies for all minted and burned assets.

\subsubsection{Limited liabilities.}
We do require that transfer transactions based on token bundles cannot lead to 
negative token quantities in accounts. However, to represent liabilities, we do 
need a notion of account with a liability, albeit one of limited lifetime.

To this end, we introduce the concept of a \emph{temporary anonymous accounts.} 
A temporary anonymous account is being identified with a hash derived from the 
transaction hash. Moreover, such an account is only available within one batch 
of transactions. In other words, just like a liability, it never gets added to 
the ledger state. As a logical consequence, a temporary anonymous account must 
have a balance of zero at the end of the batch in which it is active. This 
ensures that temporary anonymous accounts are never included in the ledger 
state.

In contrast to regular accounts in the ledger state, a temporary anonymous 
account may hold a token bundle, where one or more assets occur in a negative 
quantity. Those accounts, thus, serve the same purpose as transaction outputs 
with liabilities in \UTXOll. Hence, they can form the basis for implementing 
babel fees for an account-based ledger.

Unlike the multi-input and multi-output \UTXO\ transactions, account 
transactions typically have only one source and one destination.
However, we can use temporary anonymous accounts as a sink as well as a source 
for the currency redistributed in a batch.

\section{Liveness} \label{sec:appx:liveness}

In order to analyze the liveness property of the spot market, we consider a 
babel 
offer that is published to the network and assume that 
there is \emph{at least one party} $b_i$ (buyer) that is attracted by this 
offer. The 
interested party will then create a transaction batch $t_{xb}$ (see Section~
\ref{sec:tx_batch}) that covers the 
liability and will publish it to the network with the expectation that this 
will (eventually) be included into a block and be published in the ledger 
implemented by the blockchain. In the following analysis, we distinguish 
between 
two cases: a) The case where all buyers are acting rationally (but not 
maliciously) and b) the case 
where a percent of the buyers are controlled by a malicious adversary party.

\subsubsection{Liveness in the presence of rational players}
In this case, the aim of the players is to increase their income by collecting 
transaction fees and accepting Babel offers. Returning to our running example, 
lets assume that some other player $b_j$ is selected to produce the next block. 
Then there are only two options: a) $b_j$ is not attracted by the offer, or 
b) $b_j$ is attracted by the offer. 

In the former case, $b_j$ will ignore the 
Babel offer, but she will not ignore the transaction batch $t_{xb}$. Since 
players are acting rationally and $t_{xb}$ includes the appropriate transaction 
fees, then $b_j$ or some other block issuer following $b_j$, will eventually 
select $t_{xb}$ to be included in a block. The higher the fees the faster this 
will take place. So liveness is preserved in this case.

In the latter case, where $b_j$ is indeed attracted by the Babel offer, then 
she can \emph{front-run} and substitute $t_{xb}$ with her own transaction batch 
$t'_{xb}$
that will accept the specific Babel offer. This is a case where front-running 
transactions is a feature: it makes it feasible for block issuers to be paid in 
the tokens they prefer for their transaction processing services. However, 
liveness is also preserved in this case since the Babel offer will be accepted 
(not by $b_i$ but by $b_j$) 
and be published to the blockchain.

\subsubsection{Liveness in the presence of adversary players}

Given that the aim of Babel fees is to facilitate transaction processing on a 
blockchain, we need to consider the adversarial case, where some of the block 
producers may conspire against the use of a particular token $T$. In our scheme 
for the implementation of Babel fees, block producers are also the buyers in 
the spot market for Babel offers. Hence, they may, in addition to ignoring 
offers with token $T$, also advertise unrealistically low exchange rates for 
$T$ in $BL$ (the table of buyers-accepted rates) to trick sellers into creating 
transactions that stand no chance of 
being processed. On the other hand, we do assume that the processing of 
transactions with Babel fees, just like directly payed transaction, is 
generally in the interest of the network. Hence, honest block producers will 
advertise a rate at which they will in fact process transactions with Babel 
fees if they are offered any.

In our analysis,  we require the ratio $\frac{t}{m-t} \leq 1 - \delta$, where 
$t$ is the 
number of 
parties controlled by the adversary and $m$ is the total number of parties and 
$\delta \in (0,1)$. This is an  \emph{honest majority assumption} 
if, without loss of generality, one assumes that all parties command the same 
amount of power, cf. \cite{backbone}.
Assuming that 
this ratio also holds for the entries in $BL$ corresponding to published rates 
for a token $T$, 
then as long as the seller includes a sufficient amount of $T$ tokens, such 
that the offer has 50\% coverage, the seller's transaction will attract 
\emph{at least one} honest party. This honest party will then issue a 
transaction batch $t_{xb}$ accepting the offer.

In the following, we show that $t_{xb}$ will eventually be published on the 
ledger and thus liveness is preserved. Assume that $t_{xb}$ is issued at round 
$r$. By the \emph{chain growth} property of the chain \cite{backbone},we know 
that the chain adopted by any honest party will keep growing. After $s$ rounds 
there will be a growth of at least $\tau s$ blocks, where $\tau$ is the chain 
velocity parameter of the chain growth property. Intuitively, $\tau$ equals the 
probability that an honest party is selected as a block issuer, which in our 
case translates to $\tau = \frac{m-t}{m}$. So after $s$ rounds (starting from 
$r$) an honest party's chain will have $\frac{m-t}{t} s$ new blocks. By the 
\emph{chain quality} property of the chain \cite{backbone}, we know that the 
ratio of \emph{honest blocks} (i.e., blocks produced by honest parties) in this 
chain will be at least\footnote{Note that the number of 
	honest blocks can be less than $\frac{m-t}{t} s$, because the adversary can 
	employ attacks such as \emph{selfish mining} that can eliminate honest 
	blocks \cite{backbone}} $\frac{m-t}{t} s \mu$, 
where $\mu \in (0,1)$ is the honest block proportion parameter of the chain 
quality property. We know that if an honest block exists, then 
$t_{xb}$ will be included, or a new $t'_{xb}$ will be included instead 
accepting the 
Babel offer by the honest block issuer. So we require that $\frac{m-t}{t} s \mu 
\geq 1$, which means:
\begin{align}
	s \geq \frac{1}{\mu}\frac{t}{m-t}
\end{align}
Due to the chain growth property of the chain, this honest block that will 
include $t_{xb}$ (or $t'_{xb}$) will be buried under $k$ blocks, where $k$ is 
the number of blocks for the \emph{common prefix} property of the blockchain 
\cite{backbone}. Then it will be reported in the ledger. Therefore, even in the 
presence of an adversary controlling $t/m$ of the total hashing power, if 
honest majority 
holds and a Babel offer attracts at least one honest party, the accepted 
offer will be (eventually) published in the blockchain and thus liveness is 
preserved.

%

\section{Proof of Theorem \ref{th:optimality}}\label{sec:proof-of-optimality}

\begin{proof}
	We will prove by induction that for any feasible block of transactions $B
	\subseteq
	\{1,2,...,j\}$ corresponding to the tuple $(t_s,t_v,r,b)$, $t_s \leq S_B
	\land
	r \leq R$, list $U[j]$ will always include some tuple $(t_s',t_v',r',b')$,
	$t_s' \leq S_B \land r' \leq R$ that dominates $(t_s,t_v,r,b)$.

	For $j=1$, we have $U[1] = \{(0,0,R,0), (s_1,v_1,R-l_1,1)\}$ and the claim
	for
	any $B \subseteq \{1\}$ trivially holds.

	We assume that the claim holds for the list $U[j-1]$.

	Let $B \subseteq \{1,2,...,j\}$ be any block and
	$(t_s,t_v,r,b)$ be the corresponding tuple, $t_s \leq S_B \land
	r \leq R$. We have two options: a) $b == 0$ and b) $b == 1$. In other
	words,
	transaction $j$ is not part of $B$, or it is part of $B$.

	If $j$ is not part of
	$B$, then by induction hypothesis we know that in the list $U[j-1]$ there
	will
	be some tuple $(t_s'',t_v'',r'',b'')$ that dominates $(t_s,t_v,r,b)$.
	Remember
	that Algorithm~\ref{alg-tx-selection} first sets $U[j] \longleftarrow
	U[j-1]$
	and then removes all dominating pairs. Thus there will be some tuple in
	$U[j]$
	that dominates $(t_s'',t_v'',r'',b'')$ and then by transitivity of
	domination
	will also dominate $(t_s,t_v,r,b)$. Thus there will be some tuple in $U[j]$
	that dominates $(t_s,t_v,r,b)$.

	If $j$ is part of $B$, then we consider block $B' = B -\{j\}$. By induction
	hypothesis again, there will be some tuple $(t_s'',t_v'',r'',b'')$ in list
	$U[j-1]$ that dominates tuple
	\begin{align*}
		(\sum_{k\in B'}^{}s_k, \sum_{k \in B'}^{}v_k(B_{k-1}'), r_k, b_k)
	\end{align*}
	Which means that $t_s'' \leq \sum_{k\in B'}^{}s_k$ and $t_v'' \geq \sum_{k
	\in
		B'}^{}v_k(B_{k-1}')$. Then, Algorithm~\ref{alg-tx-selection} will add
	transaction $j$ to the tuple $(t_s'',t_v'',r'',b'')$ and add tuple $(t_s''
	+
	s_j,t_v'' + v_j(B'),r''-l_j,1)$ to the list $U[j]$. But then we have
	$t_s'' + s_j \leq \sum_{k\in B'}^{}s_k + s_j = t_s$ and
	$t_v'' + v_j(B') \geq \sum_{k \in B'}^{}v_k(B_{k-1}') + v_j(B') = t_v$.
	Thus
	there will be some tuple in $U[j]$
	that dominates $(t_s,t_v,r,b)$.
\end{proof}

\section{Proof of Theorem
\ref{th:approx-solution}}\label{sec:proof-of-approx-solution}
\begin{proof}
	Let $B$ be the block returned from Algorithm~\ref{alg-tx-selection_apprx}
	for the problem $\\\mathtt{TxSelection}(n, S_B, M, R)$. Let $B_{opt}$ be
	the
	optimal solution for this problem. We want to show
	that $utility(B)
	\geq (1-\epsilon)utility(B_{opt}) = (1-\epsilon) OPT$. Certainly $v_{omax}
	\leq OPT$, since one possible solution is to put the most
	valuable transaction in a block by itself.
	By the definition of
	$v_i'$ we know that $\mu v_i' \leq v_i \leq (\mu + 1) v_i'$, so that $\mu
	v_i' \geq v_i
	- \mu$. These inequalities along with the fact that $B$ is the optimal
	solution for the problem $\mathtt{TxSelection}(n, S_B, M', R)$ and thus
	$\sum_{i \in B}^{}v_i'(B_{i-1}) \geq \sum_{i \in B_{opt}}^{}v_i'(B_{i-1})$,
	we have the
	following:

	\begin{align*}
		utility(B) = \sum_{i \in B}^{}v_i(B_{i-1}) \\
		\geq \mu \sum_{i \in B}^{}v_i'(B_{i-1}) \\
		\geq \mu \sum_{i \in B_{opt}}^{}v_i'(B_{i-1}) \\
		\geq \sum_{i \in B_{opt}}^{}v_i(B_{i-1}) - |B_{opt}|\mu \\
		\geq \sum_{i \in B_{opt}}^{}v_i(B_{i-1}) - n\mu \\
		= \sum_{i \in B_{opt}}^{}v_i(B_{i-1}) - \epsilon v_{omax}  \\
		= utility(B_{opt}) - \epsilon v_{omax}  \\
		= OPT - \epsilon v_{omax} \\
		\geq OPT - \epsilon OPT = (1-\epsilon) OPT
	\end{align*}
\end{proof}

\fi

\end{document}